%% file: main.tex
\documentclass[a4paper]{article}

\usepackage[english]{babel}
\usepackage[utf8]{inputenc}
\usepackage{amsmath}
\usepackage{graphicx}
\usepackage[colorinlistoftodos]{todonotes}
\usepackage{url}
\usepackage{makecell}
\usepackage{rotating}
\usepackage{booktabs}
\usepackage{tabulary}
\usepackage{fullpage}
\usepackage{wasysym}
\usepackage{multirow}
\usepackage{enumitem}
\usepackage{authblk}

\title{Evaluation-as-a-Service: Overview and Outlook}
\author[1]{Allan Hanbury}
\author[2,3]{Henning M\"{u}ller}
\author[4]{Krisztian Balog}
\author[5]{Torben Brodt}
\author[6]{Gordon V. Cormack}
\author[2]{Ivan Eggel}
\author[7]{Tim Gollub}
\author[8]{Frank Hopfgartner} 
\author[3]{Jayashree Kalpathy-Cramer}
\author[9]{Noriko Kando}
\author[10]{Anastasia Krithara}
\author[6]{Jimmy Lin}
\author[11]{Simon Mercer}
\author[7]{Martin Potthast}
\affil[1]{TU Wien, Austria}
\affil[2]{University of Applied Sciences Western Switzerland, Switzerland}
\affil[3]{Athinoula A. Martinos Center for Biomedical Imaging at Massachusetts General Hospital and Harvard Medical School, USA}
\affil[4]{University of Stavanger, Norway}
\affil[5]{plista GmbH, Germany}
\affil[6]{University of Waterloo, Canada}
\affil[7]{Bauhaus-Universit\"{a}t Weimar, Germany}
\affil[8]{University of Glasgow, UK}
\affil[9]{National Institute of Informatics, Japan}
\affil[10]{National Center for Scientific Research ``Demokritos'', Greece}
\affil[11]{Independent Consultant, Seattle, USA}

\date{\today}
\begin{document}
\maketitle
\begin{abstract}
Evaluation in empirical computer science is essential to show progress and assess technologies developed. Several research domains such as information retrieval have long relied on systematic evaluation to measure progress: here, the Cranfield paradigm of creating shared test collections, defining search tasks, and collecting ground truth for these tasks has persisted up until now. In recent years, however, several new challenges have emerged that do not fit this paradigm very well: extremely large data sets, confidential data sets as found in the medical domain, and rapidly changing data sets as often encountered in industry. Also, crowdsourcing has changed the way that industry approaches problem-solving with companies now organizing challenges and handing out monetary awards to incentivize people to work on their challenges, particularly in the field of machine learning.

This white paper is based on discussions at a workshop on Evaluation-as-a-Service (EaaS). EaaS is the paradigm of not providing data sets to participants and have them work on the data locally, but keeping the data central and allowing access via Application Programming Interfaces (API), Virtual Machines (VM) or other possibilities to ship executables. The objective of this white paper are to summarize and compare the current approaches and consolidate the experiences of these approaches to outline the next steps of EaaS, particularly towards sustainable research infrastructures.

This white paper summarizes several existing approaches to EaaS and analyzes their usage scenarios and also the advantages and disadvantages.
The many factors influencing EaaS are overviewed, and the environment in terms of motivations for the various stakeholders, from funding agencies to challenge organizers, researchers and participants, to industry interested in supplying real-world problems for which they require solutions.

EaaS solves many problems of the current research environment, where data sets are often not accessible to many researchers. Executables of published tools are equally often not available making the reproducibility of results impossible. EaaS on the other hand creates reusable/citable data sets as well as available executables. Many challenges remain but such a framework for research can also foster more collaboration between researchers, potentially increasing the speed of obtaining research results. 
\end{abstract}

\tableofcontents

\section{Evaluation-as-a-Service}
\label{sec_EaaS}
%
In areas of computer science such as Machine Learning and Information Retrieval, how well the developed algorithms function is measured by running the algorithms on data having associated ground truth showing the desired outcomes, and measuring the similarity of the algorithm outputs to the ground truth. Evaluations are often organised in a structured way in order to facilitate algorithm tests and make common data available on which algorithms can be evaluated. In the Information Retrieval area, there are regular cycles of evaluation campaigns such as the Text Retrieval Conference (TREC)~\cite{VH05}, Conference and Labs of the Evaluation Forum (CLEF)~\cite{FS14}, and the NTCIR (NII Test Collection for IR Systems)~\cite{Kud10}. In the area of machine learning, the series of PASCAL Challenges \cite{PASCAL2006} from 2005--2013 are well known. Similar initiatives exist in other domains such as computer vision.

While the idea of running such evaluations is not new, the standard approach to running evaluations in the computational sciences involves distributing the data to the groups developing the systems so that they perform the computations locally and submit the results of the computations to the organisers for evaluation~\cite{SpR1975}. This is referred to as the {\em Data-to-Algorithms paradigm\/}. Participants submit the output of their software when run on a pre-published test dataset (a so-called ``run''). This approach, however, has several shortcomings, including a complete lack of reproducibility of the shared task, and the necessity to publish test datasets prematurely, albeit sans ground truth. Notwithstanding these shortcomings, the organizers of shared tasks frequently employ run submission for its minimal organizational overhead. Criticisms of this paradigm have discussed the need for continuous evlaution and not only linked to a competition and also component evaluation, which is important to better understand the performance linked to the many components of a system~\cite{HaM2010}.

Evaluation-as-a-Service adopts the {\em Algorithms-to-Data paradigm\/}, in which the data are all stored on a (central) computational infrastructure, and participants can only access the data on this infrastructure~\cite{HML2012}. This can also avoid any contact of the researchers with the test data in case of sensitive data as only the algorithms and not the developers need to see the data.

There are currently two main commercial platforms offering Data Science competitions: Kaggle\footnote{\url{http://www.kaggle.com/}} and TopCoder.\footnote{\url{http://www.topcoder.com/}} The Dream Challenges\footnote{\url{http://dreamchallenges.org/}} offers competitions to solve biomedical challenges and have started migrating towards a cloud-based solution. All of these platforms currently use the Data-to-Algorithms paradigm, the current industry standard, by requiring the participants to download the data and submit result files. None of them offer an Algorithms-to-Data approach, which is currently the only approach that allows for scaling to Big Data and using sensitive data in competitions, such as medical data sets. 

The Data-to-Algorithms approach of distributing data is often not practical, as the data may be:
\begin{itemize}
\item Huge -- In order to obtain realistic evaluation results, the evaluation should be done on realistic amounts of data. In the case of web search, this could be Petabytes of data. The current common approach of sending these data on hard disks through the postal service or via download has its limitations.
\item Non-distributable -- In many cases, it is not permitted to distribute data due to privacy, terms of service, or commercial sensitivity of the data. Privacy is the major concern for patient records. Even though the law generally permits the distribution of anonymized medical records, large-scale anonymization can only be accomplished automatically, which data owners usually do not trust. The Twitter Terms of Service forbid redistribution of tweets, while query logs are not made available for researchers after the debacle surrounding the release of the AOL search logs in~2006. Distribution of company documents for the evaluation of enterprise search would not be permitted due to the commercial sensitivity of the data.
\item Real-time -- Companies working on real-time systems, such as recommender systems, are often not interested in evaluation results obtained on static historical data, in particular if these data have to be anonymised to allow distribution, as these results are too far removed from their operative requirements. 
\end{itemize}

Even though these drawbacks are well known, all major organisers of evaluations currently adopt the Data-to-Algorithms approach. Kaggle, for example, specifically acknowledges this on their website  by stating: ``While we are sympathetic to the fact that not everyone has access to a stellar broadband connection, the plumbing needed to move data is an unavoidable part of practicing data science.''\footnote{\url{https://www.kaggle.com/wiki/ANoteOnTorrents}, visited on 5/8/2015} Through the algorithms-to-data paradigm used by the Evaluation-as-a-Service, the necessity of moving data around is removed. The data safely remain on the servers of their owners or a trusted third party:
\begin{itemize}
\item The data does not need to be downloaded by the evaluation participants.
\item It is not necessary for the data to be seen by the participants. In the case of private data, artificial data could be made available to the participants for training purposes, and the participant programs will only have access to the data when Virtual Machines (or other containers such as the lighter Docker containers) are submitted to the competition and the participants relinquish control of the Virtual Machines to the organisers.
\item The data on the servers can be updated as regularly as needed, allowing experiments to be done on real-time data. Nevertheless, care must be taken in this case to make the competition fair by ensuring that entries are compared on the same data. As VMs including programs are submitted, the option exists to run the programs on multiple sets of data as the data evolves over time.
\end{itemize}

In the computational sciences in general, little focus has been directed toward the reproducibility of experimental results, raising questions about their reliability~\cite{FS12}. There is currently work underway to counter this situation, ranging from presenting the case for open computer programs~\cite{IHG12}, through creating infrastructures to allow reproducible computational research~\cite{FS12} to considerations about the legal licensing and copyright frameworks for computational research~\cite{Sto09}. Through preservation of data, programs, and results on a central infrastructure, EaaS should lead to improved reproducibility of research results.

\begin{figure}[tb]
\begin{center}
\includegraphics[width=0.8\textwidth]{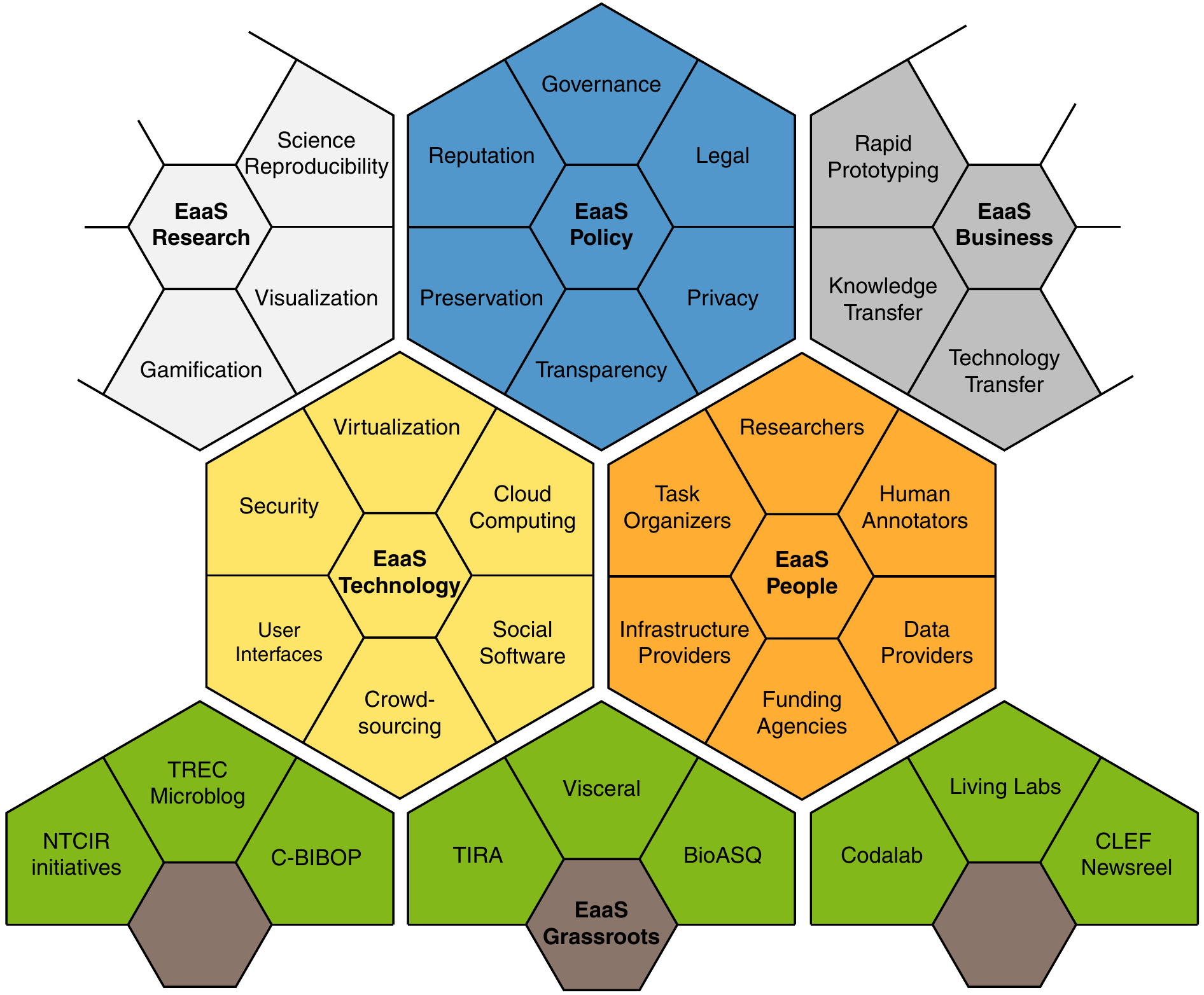}
\caption{Overview of important aspects of evaluation-as-a-service (EaaS). Aspects are grouped into five dimensions: technology, people, policy, research, and business. At the bottom of the graphic, the nine EaaS grassroot initiatives that were presented at the workshop are listed.}\label{eaas-aspects}
\end{center}
\vspace{-4ex}
\end{figure}

A number of initiatives currently implement Evaluation-as-a-Service (EaaS), either making available APIs to access the data in a controlled way, or Virtual Machines (VMs) on which systems should be deployed. In order to organize these evaluation services, various aspects need to be considered. An overview of these aspects is given in Figure~\ref{eaas-aspects}.

This White Paper is an extended version of the EaaS workshop report published as~\cite{Hopfgartner2015}. It begins by describing two success stories arising from evaluations that have used the EaaS paradigm. Then, an overview of existing EaaS initiatives is given in Section~\ref{sec_overview}. The benefits of the EaaS paradigm and the shortcomings of EaaS in its current form are covered in Sections~\ref{sec_benefits} and~\ref{sec_shortcomings}, respectively. The vision for a fully-developed EaaS approach is presented in Section~\ref{sec_vision}. Finally, aspects to consider in the next steps toward achieving this vision are covered in Section~\ref{sec_advancing}.
\section{EaaS Success Stories}
\label{sec_successes}
In this section, we detail two success stories arising from evaluations that have used the EaaS paradigm. In general the important impact of evaluations has been shown in~\cite{TJS2011,RWL2010,TLM2013}.
\subsection{More efficient indexing at the National Library of Medicine}
For the third year in a row, the National Library of Medicine (NLM) Medical Text Indexer (MTI) was used as one of the baselines for the international BioASQ Challenge. BioASQ is a series of challenges on biomedical semantic indexing and question answering with the aim of advancing the state of the art accessibility for researchers and clinicians to biomedical text (see Section~\ref{ch:BioASQ} for more details). 

The MTI indexing results are providing one of the baselines used in the ``large-scale online biomedical semantic indexing'' part of the challenge, which is designed to parallel the human indexing currently being done at NLM. The Medical Text Indexer (MTI) combines the expertise of indexers working at the NLM with natural language processing technology to curate the biomedical literature with Medical Subject Headings (MeSH) more efficiently and consistently~\cite{MDS2014}. The BioASQ Challenge provided a continuous assessment of the indexing suggestions that are automatically generated by the MTI system used in support of the MEDLINE indexing process at the NLM.

The NLM has made an announcement on the significant benefits they have from their participation in the challenge. Specifically, as mentioned by the NLM\footnote{\url{https://www.nlm.nih.gov/news/indexer_challenge.html}}, ``the benefits of participating in this community-wide evaluation for MTI were two-fold: firstly, MTI was rigorously compared to systems developed by a world-wide community of researchers and industrial teams all performing the same task; and secondly, the free exchange of the methods and ideas allowed the MTI team to incorporate the best practices explored by the participating teams. Incorporating some of these approaches into the MTI workflow in 2013--2014 improved the accuracy of MTI indexing suggestions by $4.5\%$''.


%
%

%
%
%
\subsection{Fostering Open Innovation through Living Labs}
The key to the success of commercial information retrieval systems is their ability to efficiently assist users in satisfying their information need. Successful recommender systems have to suggest relevant items that might spark the users' interest. While setting up out-of-the-box retrieval or recommender systems is relatively easy, experience shows that adapting or fine-tuning algorithms to individual use cases requires significant resources. 

More precisely, companies require talented developers, as well as knowledgeable engineers who have both a good overview of state-of-the-art techniques and the capability to develop innovative solutions to improve their systems' performance. This, however, can be rather challenging for companies, especially if they are small and do not have the required funds available. The idea of EaaS can serve here as a possibility to bridge this innovation blocker. In particular, it is a type of Open Innovation, defined as ``Open Innovation is a paradigm that assumes that firms can and should use external ideas as well as internal ideas, and internal and external paths to market, as they look to advance their technology'' \cite{CVW06}.

An example EaaS is implemented in NewsREEL, the first initiative that allows the evaluation of information access techniques in a living laboratory setting. NewsREEL addresses the challenge of news recommendation in real-time. The initiative was initially funded via a national knowledge transfer program that aims to foster collaboration between SME's and academia. Within the project, a commercial provider of news and ad recommendations developed a platform that allowed the academic project partner to develop recommendation algorithms and have them embedded in the live system of the provider. By providing statistics about the efficiency of the algorithms, the platform efficiently allowed the academic partner to perform A/B tests in a living lab. From the very beginning, this platform was open to anyone who wants to develop and benchmark news recommendation algorithms. By organising NewsREEL as a campaign-style lab of CLEF, participants were motivated to report their efforts at a scientific conference. 

For the company that provided the infrastructure and API of NewsREEL, the advantages are manifold. First of all, it allowed them to raise awareness of their company and their use case. In fact, following the start of NewsREEL, the company received a number of job applications that directly referred to the evaluation campaign. Moreover, by interacting with the participants, they can learn of innovative ideas on how to address their recommendation task. Finally, by allowing participants to benchmark the performance of their ideas in a live system, the company can save expensive development time. 

For academia, advantages are manifold as well. First of all, it allows researchers to gain experience in developing innovative techniques and have them tested under real live conditions. Moreover, it provides access to the infrastructure and large user base of a commercial service provider, hence reducing the gap between academic and industry research.
\section{Existing EaaS Initiatives}
\label{sec_overview}
This section begins by providing details on currently existing evaluation initiatives that use the Evaluation-as-a-Service paradigm. It then discusses some of the management systems used in facilitating Eaas. Finally, a comparison between the initiatives based on key characteristics of EaaS is provided.
\subsection{Description of Initiatives}
In what follows, more information on the existing initiatives is given.
\subsubsection{TREC Microblog Task}
The TREC Microblog tracks began in 2011 to explore search tasks and evaluation methodologies for information seeking behaviors in microblogging environments such as Twitter. TREC 2015 marks the fifth iteration of the track. For the past four years, the core task has been temporally-anchored ad hoc retrieval, where the putative user model is as follows:\ ``At time $T$, give me the most relevant tweets about an information need expressed as query $Q$.''  Since its inception, the track has had to contend with challenges related to data distribution, since Twitter's terms of service prohibit redistribution of tweets. For TREC 2011 and 2012~\cite{Ounis_etal_TREC2011}, the track organizers devised a solution whereby  the {\it ids} of the tweets were distributed, rather than the tweets themselves. Given these ids and a downloader program (also developed by the track organizers), a participant could ``recreate'' the collection~\cite{McCreadie_etal_SIGIR2012}. This approach adequately addressed the no-redistribution issue, but was not scalable as participants in the end had to recreate the collection locally. TREC 2013~\cite{Lin_Efron_TREC2013} implemented an entirely different solution, which was to provide an API through which participants could complete the evaluation task. That is, the organizers gathered a collection of tweets centrally, but all access to the collection was mediated through the API, such that the participants could not directly interact with the raw collection. The search API itself was built using Thrift\footnote{\url{http://thrift.apache.org/}} and the Lucene search engine,\footnote{\url{http://lucene.apache.org/}} which are both widely-adopted open-source tools. A nice side-effect of the API approach is that common infrastructure promotes reproducibility~\cite{Rao_etal_ECIR2015} and sharing of open-source software components.
\subsubsection{BioASQ}
\label{ch:BioASQ}
BioASQ aims to push research towards highly precise biomedical information access systems by establishing a series of challenges in which systems from teams around the world compete~\cite{tsatsaronis2015overview}. BioASQ provides data, software and the evaluation infrastructure for the challenge. By these means, the project ensures that the biomedical experts of the future can rely on software tools to identify, process and present the fragments of the huge space of biomedical resources that address their personal questions. BioASQ comprises two tasks. In Task~A systems are required to automatically assign MeSH (Medical Subject Headings) terms to biomedical articles, thus assisting the indexing of biomedical literature. This task uses the EaaS approach to include participating systems directly in the indexing process of the National Library of Medicine (NLM) --- Systems participating in the task are given newly published MEDLINE articles, before the NLM curators have assigned MeSH terms to them. The systems assign MeSH terms to the documents, which are then compared against the terms assigned by the NLM curators. Task~B focuses on obtaining precise and comprehensible answers to biomedical questions. The systems that participate in Task~B are given English questions written by biomedical experts that reflect real-life information needs. For each question, the systems are required to return relevant articles, snippets of the articles, concepts from designated ontologies, RDF triples from Linked Life Data, an `exact' answer (e.g., a disease or symptom), and a paragraph-sized summary answer \cite{tsatsaronis2015overview}.
\subsubsection{VISCERAL}
\begin{figure}
\begin{center}
\includegraphics[width=0.9\textwidth]{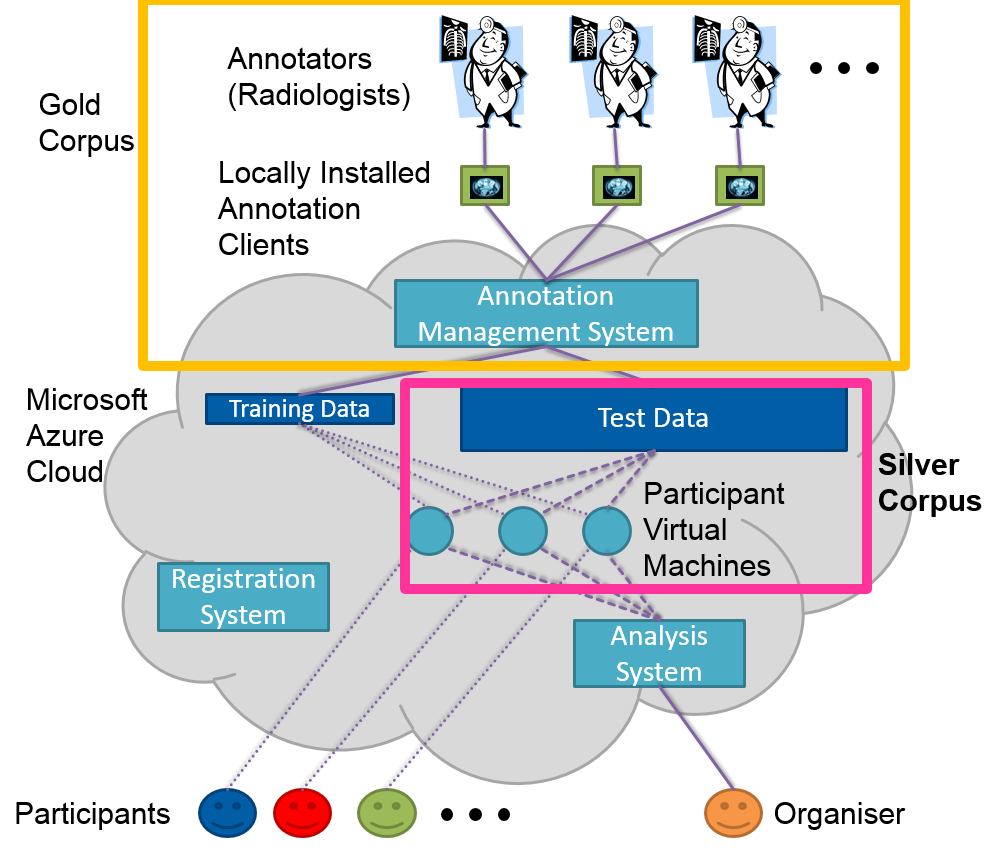}
\caption{Overview of the VISCERAL cloud-based approach.}\label{fig_VISCERAL}
\end{center}
\end{figure}

The FP7 project VISCERAL\footnote{\url{http://visceral.eu/}} organized a series of benchmarks on the processing of large-scale 3D radiology images~\cite{LMM2012}. The tasks include the segmentation of images, the detection of lesions in the images and the retrieval of similar cases including images and semantic terms as queries.
VISCERAL made use of an innovative cloud-based evaluation approach, illustrated in Figure~\ref{fig_VISCERAL}, where all data are stored in the cloud. Participants in the tasks get Virtual Machines (VMs) to install their software and access to training data via the cloud. For the test phase the virtual machines are blocked for the participants and the organizers take over the VMs and run the executables connecting the VM to a different storage with the test data. The use of the cloud also facilitates the creation of ground truth. A {\em Gold Corpus} of manually segmented organs is created by radiologists, with the process managed by an Annotation Management System. This system sends tickets to radiologists hired as annotators with instructions on the organ to segment, and tracks the annotation progress, allowing the process to run efficiently. It also manages a quality control process by which the manual segmentations are controlled. The executables submitted by participants are also used in collaboration with the participants to run the algorithms on more non-annotated data sets with a goal to use label fusion and create more ground truth by fusing the output of all participant approaches. The ground truth created in this way is called the {\em Silver Corpus\/}~\cite{KDJ2015}.

\subsubsection{C-BIBOP}
Cloud-based Image Biomarker Optimization Platform (C-BIBOP)\footnote{\url{http://cbibop.org/}} is being developed as a technical resource for the cancer research community to support the development and assessment of quantitative imaging biomarkers. Lesion segmentation is a critical step in the development and use of imaging biomarkers in cancer. Another task that is organized as part of C-BIBOP requires the analysis of Magnetic Resonance Imaging (MRI) to identify biomarkers that best correlate with clinical outcomes. C-BIBOP is being developed to support reproducible science by enabling researchers to compare the performance of their image analysis algorithms that are co-located with large medical imaging datasets. The size of the datasets as well as the concerns about the sensitive nature of the data has highlighted the need for cloud-based solutions. Evaluation-as-a-Service allows the challenge organizers to customize the evaluation methods for the clinical questions being addressed. Currently, C-BIBOP is built on the CodaLab platform and plans to integrate key aspects from the VISCERAL project.
\subsubsection{CLEF NewsREEL\label{CLEFNEWSREEL}}
\begin{figure}
\begin{center}
\includegraphics[width=0.8\textwidth]{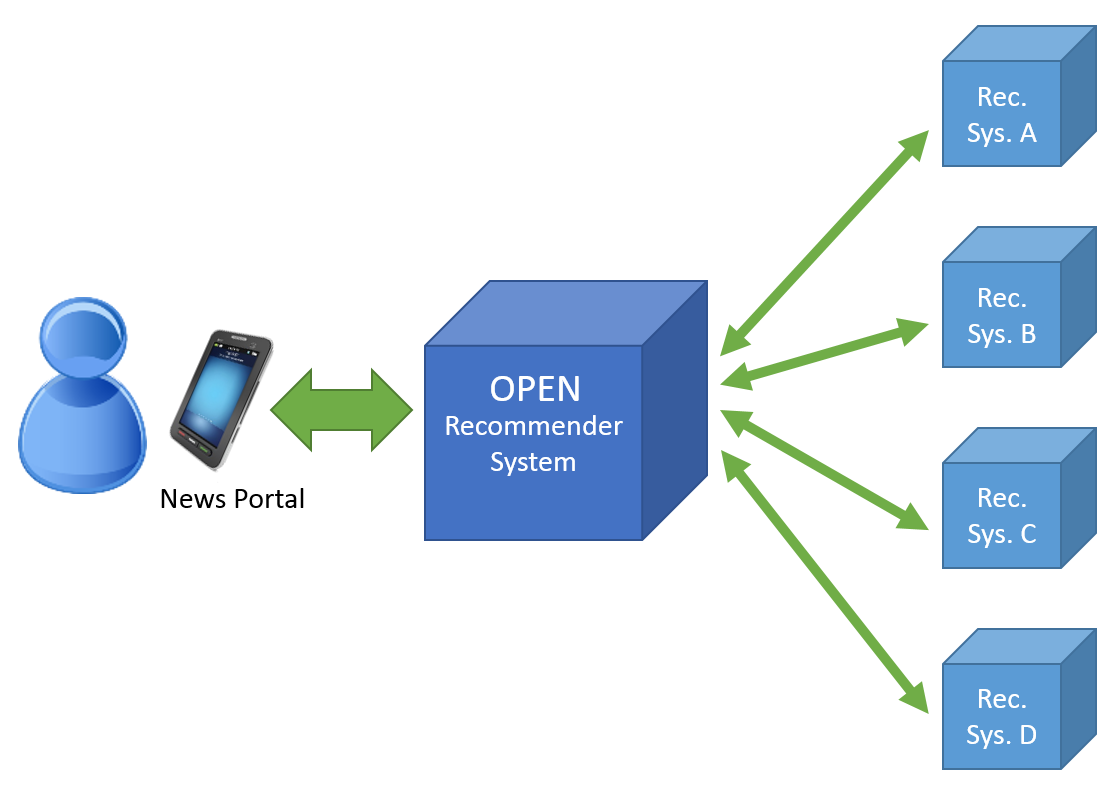}
\caption{Overview of the CLEF NewsREEL approach.}\label{fig_CLEF_NewsREEL}
\end{center}
\end{figure}

The News REcommendation Evaluation Lab (NewsREEL)\footnote{\url{http://clef-newsreel.org/}} is a campaign-style evaluation lab of CLEF. It implements the idea of \textit{living laboratories} where researchers gain access to the resources of a company to evaluate different information access techniques using A/B testing~\cite{Hopfgartner:2014}. The infrastructure is provided by plista GmbH, a company that provides a recommendation service for online publishers. Whenever a user requests an article from one of their customers' web portals, plista recommends similar articles that the user might be interested in. In NewsREEL, plista outsourced this recommendation task for a selected subset of their customers to interested researchers: Participants are asked to provide recommendations in real-time for actual users, i.e., the list of related articles is not determined by plista, but by the participating research teams. The communication between the participants and plista, as well as the monitoring and evaluation is handled by the Open Recommendation Platform (ORP)~\cite{Brodt:2014} (see Section~\ref{sec:ORP} for further details). ORP serves as a Web Service that is constantly sending users' requests for articles, as well as informs about new articles being added by the publisher, or existing articles being updated. The platform allows participants to register various recommendation algorithms in parallel and benchmark their performance over a longer period of time, as illustrated in Figure~\ref{fig_CLEF_NewsREEL}. 

In the context of CLEF, four evaluation periods of two weeks duration each are defined during which the performance of participants' algorithms are measured and compared to a baseline run. The selected performance metric is the click-through rate, i.e., the ratio between the number of requested recommendations and the number of recommendations that users clicked on.
\subsubsection{CLEF Living Labs for Information Retrieval}
Living Labs for Information Retrieval (LL4IR)\footnote{\url{http://living-labs.net/}} is an effort similar to NewsREEL, also running as a CLEF lab, but focusing on retrieval as opposed to recommendation. 
LL4IR provides a benchmarking platform where researchers can gain access to privileged commercial data (click and query logs) and can evaluate their ranking systems in a live setting, with real users, in their natural task environments. The first edition of the lab focuses on three specific use-cases: product search (on an e-commerce site), local domain search (on a university' website), and web search (through a major commercial web search engine). A key idea to removing the harsh requirement of providing rankings in real-time for query requests is to focus on head queries~\cite{Balog2014Head}.  Participants can produce rankings for each query offline and upload these to the commercial provider.  The commercial provider then interleaves a given participant's ranked list with their own ranking, and presents the user with the interleaved result list.  Finally, feedback is made available to participants to facilitate improved offline ranking generation.  Data exchange between live systems and participants is orchestrated by a web-based API.
\subsubsection{PAN Shared Task Series on Digital Text Forensics}
PAN is a networking initiative for digital text forensics~\cite{stein:2015m}\footnote{\url{http://pan.webis.de/}}
where researchers and practitioners study technologies that analyze texts with regard to originality, authorship, and trustworthiness. The practical importance of such technologies is obvious for law enforcement and marketing, yet the general public needs to be aware of their capabilities as well to make informed decisions about them. This is particularly true since almost all of these technologies are still in their infancy and active research is required to push them forward. PAN therefore focuses on the evaluation of selected tasks from the digital text forensics in order to develop large-scale, standardized benchmarks, and to assess the state of the art. An important goal of PAN over the past years has been to establish shared task competitions that are reproducible, so that future evaluations within and without PAN can be done in comparison to the state of the art. To attain true reproducibility in a shared task competition, however, it is necessary to allow for exchanging all of its building blocks, including software, data, and performance measures at any time. Neither of them can be assumed fixed forever, so that once someone proposes, for instance, a new dataset, it should be possible to re-evaluate all existing software on this new dataset. This insight informed PAN's move to adopt the Evaluation-as-a-Service paradigm for all of its shared tasks since~2012. PAN employs the TIRA experimentation platform (see Section~\ref{tira}), where software, datasets, and performance measures can be deployed in the cloud, and where the software solving a given task can be remotely executed. This way, PAN has assembled the largest collection to date of more than 150~pieces of digital text forensics software from researchers all over the world.
\subsubsection{CoNLL Shared Task}
The Conference on Natural Language Learning (CoNLL)\footnote{http://ifarm.nl/signll/conll/}
organized by the ACL Special Interest Group on Natural Language Learning (SIGNLL)\footnote{http://ifarm.nl/signll/}
has been an early adopter of shared task evaluations in the natural language processing community. Since~1999, the conference has organized annual shared tasks on various important problems of natural language learning as a regular part of the conference program. Some of the evaluation resources that have been developed for these shared tasks have become standard benchmarks and are widely used today. However, the software that has been developed for these shared tasks throughout the years has not been collected by the shared task organizers but remains with their participants. In time, the chances of being able to obtain certain pieces of software decrease rapidly, since the researchers responsible move on in their careers and may no longer be available. This has been recognized as a major limitation to the reproducibility of CoNLL's shared tasks, so that SIGNLL has decided to adopt the emerging Evaluation-as-a-Service paradigm for the 2015 shared task~\cite{xue:2015}.%
\footnote{http://www.cs.brandeis.edu/~clp/conll15st/}
CoNLL employs the TIRA experimentation platform (see Section~\ref{tira}), where the evaluation datasets and performance measures for the shared task have been deployed, and participants have been invited to deploy their software into TIRA's virtual machines. Altogether, 16 teams have submitted software to the shared task, demonstrating the transition of CoNLL's shared task to the Evaluation-as-a-Service paradigm as implemented by TIRA did not cause participation rates to decrease.
\subsubsection{TREC Total Recall Track}
The principal purpose of the Total Recall Track 2015 was to evaluate, through
a controlled simulation, methods to achieve very high recall \textendash{}
as close as practicable to 100\% \textendash{} with a human assessor
in the loop. Motivating application domains include legal eDiscovery~\cite{grossman2014comments}, systematic reviews for meta-analysis in evidence-based
medicine~\cite{lefebvre2008searching}, and the creation of fully labeled
test collections for information retrieval evaluation~\cite{cormacklynam05}. A secondary
\textendash{} but important \textendash{} purpose was to develop a
sandboxed virtual test environment within which information retrieval
systems may be tested while preventing the disclosure of sensitive
test data to participants. At the same time, the test environment
operates as a black box, affording participants confidence that their
proprietary systems cannot easily be reverse engineered.

The task to be solved was:
\begin{quote}
Given a topic description (like those used for ad-hoc and web tasks),
identify the documents in a corpus, one at a time, such that, as nearly
as possible, all relevant documents are identified before all non-relevant
documents. Immediately after each document is identified, its ground-truth
relevance or non-relevance is disclosed.
\end{quote}
Datasets, topics, and automated relevance assessments were all provided
by a Web server supplied by the Total Recall Track. Participants were
required to implement either a fully automated or semi-automated ({}``manual'')
process to download the datasets and topics, and submit documents
for assessment to the Web server, which rendered a relevance assessment
for each submitted document in real time. Thus, participants were
tasked with identifying documents for review, while the Web server
simulated the role of a human-in-the-loop assessor. Rank-based and
set-based evaluation measures were calculated based on the order in
which documents were presented to the Web server for assessment, as
well as the set of documents that had been presented to the Web server
at the time the participant declared that a {}``reasonable'' result
had been achieved. Particular emphasis was placed on achieving high
recall while reviewing the minimum possible number of documents.
\subsubsection{MIREX}
The Music Information Retrieval Evaluation eXchange (MIREX) is the community-based framework for the formal evaluation of Music Information Retrieval (MIR) systems and algorithms~\cite{Dow2008}, which has been running annualy since 2005. This evaluation campaign has the difficulty that distributing the music recordings on which the evaluation tasks are run is not permitted due to copyright. To compensate for this limitation, participants are required to upload executable files that carry out the tasks that are evaluated, and these executables are run on a single repository of music files. An online submission system supports the submission process. Running the executables on the repository of music files is done manually by the principal organiser of MIREX --- to facilitate this, participants must adhere to a specification for calling the executable, and must provide details on software/architecture dependencies and other configuration details. 








%
\subsection{EaaS Management Systems}
This section contains descriptions of a selection of EaaS management software tools that are either available as a running service and have been used by multiple evaluation campaigns or challenges, or for which the source code has been made available. 
\subsubsection{TIRA}
\label{tira}
The TIRA experimentation platform is a web service that supports organizers of shared tasks in computer science to accept the submission of executable software~\cite{stein:2012k}.
TIRA\footnote{\url{http://www.tira.io/}} automates software submission to a point at which it imposes no significant overhead on organizers and participants alike. From the start, TIRA has been in active use: since 2012, TIRA is employed for the PAN shared task series on digital text forensics~\cite{stein:2014j}, and as of 2015, TIRA hosts the annual shared task of the CoNLL~conference~\cite{xue:2015}. TIRA's technology stack relies primarily on a combination of low-level (LXC, Docker) and high-level (hypervisor) virtualization technology, server-side control software, and a web front end that allow for the remote management of shared tasks. TIRA distributes virtual machines across a number of TIRA hosts, which are remote-controlled by a master server. Every virtual machine is accessible from the outside by participants via SSH and remote desktop, and both Linux and Windows are supported as guest operating systems. This allows for a variety of development environments, so that participants in a shared task can directly work as they usually would. TIRA further hosts the datasets used in a shared task, split into training datasets and test datasets. The former are publicly visible to participants, including ground truth data, whereas the latter are accessible only to participant software in a secure execution environment that protects the test datasets from leaking to participants. Before executing the software on a test dataset, TIRA clones its virtual machine into the secure execution environment, where Internet access is disabled. After the software successfully executed on the test dataset, its output is copied, whereas the cloned virtual machine is deleted to prevent any potentially private files on its virtual harddisk from exiting the execution environment. In this way, participants in a shared task can run their software on the shared task's test datasets, whereas its organizers need not worry about the data leaking. TIRA also enables the use of proprietary and sensitive data as evaluation data. Finally, TIRA hosts a special purpose virtual machine for each shared task, where the organizer deploys software for performance measurement. The output of participant software that was executed on a training dataset or a test dataset is fed directly into the performance measurement software at the click of a button. The results are displayed on a dedicated web page for the shared task on TIRA's web front end.
\subsubsection{VISCERAL Registration System}
The VISCERAL Registration System is the software used to manage the VISCERAL benchmarks. The code is available.\footnote{\url{https://github.com/Visceral-Project/registration-system}} The system is currently closely tied to the Azure cloud. It supports participant registration, provision of benchmark documentation and files to participants, management of VM assignment, and almost fully automated VM submission and result evaluation for image segmentation benchmarks. The metrics calculated for each submission are available to the participant that submitted, and a participant can choose to make the results publicly visible on a leaderboard.
\subsubsection{Open Recommendation Platform (ORP)\label{sec:ORP}}
The key component of the News Recommendation Evaluation Lab that is introduced in Section~\ref{CLEFNEWSREEL} is the Open Recommender Platform (ORP) \cite{Brodt:2014}. 
After registering an account on the platform, participants need to provide a server address (and port number) and activate their account. ORP then starts broadcasting item updates, event notifications and recommendation requests. Event notifications are the actual user interactions, i.e., user visits, referred to as impressions, to one of the news portals that rely on the plista service, or clicks to one of the recommended articles. The item updates include information about the creation of new pages on the content providers' server and it allows participants to provide content-based recommendations. Expected responses to the recommendation requests are related news articles from the same content provider, which are then provided as recommendations to the visitors of the page. Requests are send in the form of HTTP POST requests. JSON is used for data encoding. 

One of the main requirements of this scenario is that recommendations have to be provided in almost real-time. Considering that a constant data stream~\cite{Kille:2013} is exchanged between ORP and the participants' server, network latency becomes an actual issue since it reduces the amount of time remaining to compute recommendations. In order to avoid this time loss, plista allows participants to run their algorithms on VMs in their data center.

ORP provides a Web interface that consists of five different tabs, namely dashboard, statistics, debugging, leaderboard, and documentation. The dashboard allows users to set up their server and activate individual algorithms. The statistics page visualizes the performance of the registered algorithms. The leaderboard page shows the overall performance of all teams that are currently participating in the challenge. 
\subsubsection{CodaLab}
The CodaLab platform\footnote{\url{http://www.codalab.org}} is an ongoing open-source development project with the goal of encouraging researchers to share and interact with datasets and algorithms through the medium of online scientific competitions. Written in Python, CodaLab both supports the standard academic model of competition in which participants download a common dataset, execute their algorithm locally and upload their results, but at the discretion of the competition owner it can also use the Microsoft Azure cloud to provide a standardized execution environment. 

Any user can create a competition, defining multiple phases and automating the evaluation criteria needed to pass from one phase to the next. This may be done using either the editor provided or by uploading an appropriately-structured file --- extensive documentation is available in the GitHub repository\footnote{\url{https://github.com/codalab/codalab/wiki/}}. While the medical image analysis community were early adopters of CodaLab, the system offers sufficient flexibility to be useful to the scientific community in general and is now used more widely, e.g., for some recent challenges on ChaLearn.\footnote{\url{http://www.chalearn.org/}}
\subsubsection{OpenML}
OpenML\footnote{\url{http://www.openml.org}} is a platform that allows machine learning researchers to share data, code and results (e.g., models, predictions, and evaluations) \cite{VRBT2013}. The types of objects that OpenML currently handles are data, tasks, flows and runs. {\em Data} can be uploaded to the platform or linked to by a URL. {\em Tasks} describe what should be done with a data set, and include additional information such as training/test splits and what needs to be returned. Tasks can be of various types such as machine learning, clustering and regression. {\em Flows} are algorithms, workflows, or scripts for solving tasks, and {\em Runs} are applications of flows on tasks. Runs contain all information necessary to make the experiment reproducible, including data, flows, and parameter settings. All objects are searchable on the OpenML platform.
\subsubsection{TREC Total Recall Management System}
The TREC 2015 Total Recall Track used three modes of participation:
{} ``Practice'' participation, {}``At Home'' participation, and {}``Sandbox'' participation. Practice and At Home participation
was done using the open Web: participants ran their own systems and
connected to the Web server at a public address. The Practice collections
were available for several weeks prior to the At Home collections;
the At Home collections were available for official runs throughout
July and August 2015 (and continue to be availalbe for unofficial
runs).

Sandbox runs were conducted entirely on a Web-isolated platform hosting
the data collection. To participate in the Sandbox task, participants
were required to encapsulate -- as a VirtualBox virtual machine --
a fully autonomous solution that would contact the Web server and
conduct the task without human intervention. The only feedback available
to participants consisted of summary evaluation measures showing the
number of relevant documents identified, as a function of the total
number of documents identified to the Web server for review.

To aid participants in the Practice, At Home, and Sandbox tasks, as
well as to provide a baseline for comparison, a Baseline Model Implementation
(BMI) was made available to participants.\footnote{\url{http://plg.uwaterloo.ca/~gvcormac/trecvm/}} BMI was run on all the collections,
and summary results were supplied to participants for their own runs,
as well as the BMI runs.
\subsection{Comparison between Initiatives}
The EaaS initiatives are compared using the following characteristics (see Table~\ref{table-eaas-initiative-comparison} for an overview):
\begin{description}
\item[Software:] If one of the available EaaS management systems is used, or if dedicated software was written for the initiative.
\item[Data:] If the data is a static collection or if the evaluation is run on dynamic (real-time) data.
\item[Data Access:] How the participants get access to the data. Possibilities include downloading the data, interacting with an API, or accessing data stored on the cloud via a Virtual Machine (VM) on the cloud.
\item[Submission:] How the results are submitted. Possibilities are by uploading result files in a specified format, by interacting with an API, or by submitting code installed on a VM.
\item[Continuous:] If the system allows for continuous evaluation with results submitted at any time, or if there is a fixed deadline for result submission.
\item[Automation:] The degree of automation of the result processing, where result processing includes collecting participant submissions, analysing them, and making the results available to participants. Possibilities are {\em little\/}, meaning that the organisers conduct the result processing almost completely manually, {\em partly\/}, meaning that significant parts of the result processing are automated, and {\em fully\/}, meaning that only very minor interactions are needed from the organisers (on the order of interacting with the system by clicking buttons to start processes).
\item[Result Interaction:] The type of interaction that the EaaS system allows with the submitted results, e.g. visual analytics functionalities for result comparison.
\item[Technical Support:] If and how technical support is provided.
\end{description}

%
\input{table-eaas-initiative-comparison}
\section{Benefits of EaaS}
\label{sec_benefits}
The goal of this section is to highlight what participants and other stakeholders, such as campaign organizers or companies crowdsourcing the technology development by proposing tasks, get as benefits and why they participate in these types of events. Funding agencies and science as a whole can get benefit from the EaaS paradigm as all projects funded can become comparable and data are not limited to a small group that can use them but can be shared virtually for the analysis. This can lead to generally better science that is more reproducible and where more time can be taken for the large data creation as more people can work together instead of creating many small datasets. Below we consider the benefits for participants, companies proposing tasks, organisers and science as a whole.

The participant benefits usually relate to the following points:
\begin{itemize}
\item access to annotated data sets and a clear evaluation scenario, making it quick and easy to publish if the results are good;
\item access to data sets that would be too big to be shared and that a single research group or a small group could not assemble and treat;
\item access to sensitive data such as medical data but also in other domains (copyrighted music, enterprise search data). Without EaaS the companies would likely not share the data but maybe work on it in-house only, such as large search engine companies currently do with their log files;
\item get a comparison to strong baselines, so other techniques and algorithms do not need to be reimplemented and then optimized; this has the reverse risk that it can make one's own results look less positive than comparing to a low baseline~\cite{AMW2009a};
\item get impact via publications, mainly by reusing the data after the end of competitions for further publications;
\item if sharing of components is done, then this could also give more visibility, citations and reputation but this is currently not very often the case;
\item advertisement via demos that are dissemination channels of own techniques;
\item workshops to discuss with people working on the same data to get ideas on new approaches and avoid mistakes others have done but not published, as publications of negative results are rare;
\item access to a broader range of challenges and testing own tools on the data best adapted for them;
\item potentially better contacts to business partners if the challenges are proposed by a company for example, this could also lead to job offers for graduate students.
\end{itemize}
It is clear that the motivation of researchers can be intrinsic, so a student looking for a good scientific reputation through using good techniques. It can be extrinsic as well, for example winning prize money at a competition. For senior researchers, winning a competition can also lead to an easier access to public funding for the techniques or also industry funding if winning an important competition.

The potential advantages of EaaS for a company that organizes tasks to crowdsource part of their innovation processes are:
\begin{itemize}
\item tapping into the skills of external data analytics experts to solve a data analytics problem for the company with a limited amount of funding available;
\item obtaining a solution that is potentially much better than would have been produced internally as many more tools and algorithms can be tested and compared;
\item identifying and getting access to new talent for hiring and already knowing about their skills and qualities beforehand;
\item if a Virtual Machine containing software or a Docker container is submitted, the company can also test the software on other data (as long as this is specified in the participation agreement) and can test generalizability.
\end{itemize}
Organizers of evaluation campaigns also have several benefits:
\begin{itemize}
\item getting a possibility to shape the task of many scientists, so shaping research directions and getting influence in this;
\item getting potentially many citations if data sets are reused and could become the standard for a specific sub field;
\item getting the possibility to position themselves as leaders in the field if people use the data and scenarios provided;
\item getting access to the best performing techniques and a clear idea of techniques, their performance and their stability based on the work of participants, which can also lead to new ideas or projects based on real data;
\end{itemize}
Science as a whole can benefit from such an approach in the following ways:
\begin{itemize}
\item efficiency of research will increase as data sets are reused. By working together larger data sets with more statistical power can potentially be obtained and work on creating small and potentially meaningless data sets is no longer necessary;
\item community building around a problem can help everyone to get in contact with people working on similar problems and data and thus foster collaboration and collaborative problem solving, potentially also combining code and reusing existing tools that work well;
\item reproducibility of papers is extremely high if data and evaluation scenarios are available in addition to executable code and all can be cited and rerun if necessary, also on new data or extended data sets;
\item entry barriers to these domains for new persons such as PhD students are much lower if tools and data can be reused and all energy is spent on improving tools and new approaches based on strong baselines and not getting an evaluation scenario organized.
\end{itemize}
Many of these advantages are also strong incentives for funding organizations that can favor projects participating in such tasks or even organizing tasks or providing data. In addition to performance comparisons of algorithms, run time of algorithms can be measured if evaluated on the same infrastructure, so effectiveness and efficiency are used for ranking approaches.

Some of these points are definitely valid for all types of evaluation campaigns and data sharing but some are closely linked to the paradigm of EaaS, as the code is available and can be reused, for example, for creating a silver corpus on new data~\cite{KDJ2015}.
Having running code in addition to data and published papers can also help to speed up the direct transfer of technology to industry.
\section{Shortcomings of Existing EaaS Initiatives}
\label{sec_shortcomings}
%
Besides all the advantages mentioned just before, there are also a few entry barriers and problems for participants in competitions, particularly if EaaS is used, so no simple download of data but the requirement to provide executables: 
\begin{itemize}
\item participants need to reinstall the full software if virtual machines are used, so this is harder than running tools locally; even though using Docker could reduce this problem as software can be moved more easily between local machines and the evaluation infrastructure;
\item if participants are of big and well known groups in the field then poor performance can hurt their reputation and for this reason some groups only participate in challenges that they feel have a very high chance to have very good results;
\item some software tools frequently used for research such as MatLab usually require access to license servers so if the evaluations are run in a totally closed environment this can mean that the software might not run properly or additional adaptations are necessary;
\item participants in VISCERAL mentioned to feel a loss of control if they do not feel the data (or have it locally) as this is what most people are used to and it allows to quickly check the data visually and the first results; this obviously does not scale to big data;
\item errors on the test data could be different from errors on the training data and this might only become visible after the end of a competition run and limit performance for potentially good techniques, which is also related to the control loss;
\item sustainability is a problem as well for cloud or local infrastructure and installing everything for one single run of a competition might not be worth the effort but if it remains reusable then the effort might be worthwhile;
\item VMs may not be sufficiently powerful, or do not have specific hardware such as GPUs available that some participants need for their algorithms to run well. For very large data sets the software tools are increasingly adapted to specific hardware for efficiency reasons --- this becomes a problem if standard hardware in the cloud is being used and emulating specific hardware does not always work well.
\end{itemize}
Also the campaign organisers have to deal with potential problems:
\begin{itemize}
\item when work is done on confidential data the organizers need to make all security provisions and they may be held responsible for any shortcoming in the security infrastructure;
\item manual feedback for participants is costly (mails etc.), but it can be necessary if problems occur that could not be visible on the training data to make sure that no participant feels disadvantaged in the evaluation, particularly if price money is involved;
\item legal questions can arise that are different from standard benchmarks, for example, when reusing the code of participants for other tasks, and it can mean to take risks for participants; if companies want to make sure that their proprietary code is safe then also it is the organizer who needs to assure this;
\item participants in existing cloud-based campaigns left VMs running without doing anything in them, which causes costs and is difficult to prevent entirely.
\item as funding often ends after projects but is necessary to keep data sets alive and allow for evaluation if cloud computing is used it is also necessary to think about other financial models than project funding but something long term and sustainable that might likely be in the way of a public-private partnership;
\end{itemize}
There are risks that companies and funding agencies need to take into account:
\begin{itemize}
\item losing reputation for companies if results are not good is a real risk and for this reason some possibilities to remove runs maybe be needed;
\item if a funding agency supports an evaluation campaign it is important that validity is assured, for example, by really taking the best and meaningful performance measures; also statistical power needs to be checked as otherwise there is a risk that the results will in the end not mean much and the best and worst groups are extremely close together;
\item in general, the concentration on competitions can limit the number of new techniques, as there is a higher incentive to use small modifications of existing techniques than to develop something really new with maybe a higher potential but also a higher risk that in the beginning the performance is limited.
\end{itemize}
\section{EaaS Vision}
\label{sec_vision}
Once it is fully developed, the EaaS paradigm should provide advantages for both academia and industry, with academia getting access to interesting data and challenges, and industry getting access to results that could improve the offerings to their clients. To illustrate the advantages obtained by academia and industry, we first present two scenarios illustrating how EaaS could be used once it is fully developed:
\begin{quotation}
\noindent {\bf Scenario 1:} {\em Company X} provides a blog entry recommendation service based on a user profile and a user click history. {\em Company X} wishes to improve its recommendation algorithms, and decides to make this challenge open to all through the EaaS paradigm, thereby increasing the size of the pool of highly-qualified people from which the solution can come. Participants submit their proposed recommender systems as executables installed on Virtual Machines, where the parameters such as maximum response time are strictly specified. Due to the use of a standardised VM exchange format, the participation overhead is reduced, as the same VM images can be submitted to participate in any EaaS. Upon submission of the VMs, standardised tests are automatically run to ensure that the systems satisfy the specified parameters, and any shortcomings are reported back to the participant. Once a system satisfies all parameters, it is randomly assigned requests for recommendation, and evaluated based on the clicks by end users of the {\em Company X} recommendation service on the links returned. A well-designed experimental protocol ensures that links suggested by the submitted recommender systems are shown often enough to obtain statistically significant results for all participants, and participants are sent results in the form of performance metric values linked to a permanent identifier that are ready to be inserted directly into a publication. {\em Company X} gets information on how well the performance of their recommender system compares to the state-of-the-art, and can contact the teams having the best performance to discuss potential technology transfer.
\end{quotation}

\begin{quotation}
\noindent {\bf Scenario 2:} {\em Company Y}, a pharmaceutical company, wishes to make drug development for {\em Disease Z} more efficient and cost-effective. It identifies that there are two main components to doing this, better extraction of key information from the biomedical literature, and better prediction of the outcomes of combining various ingredients. {\em Company Y} wishes to get the solution from the largest possible pool of experts, and therefore opens the challenge as EaaS. The two parts of solving the challenge are heavily interdependent (as prediction is influenced by the available facts), but require different skill sets on the part of the solution providers, so participants can select to participate in either extraction or prediction. A huge collection of biomedical literature and of facts that are already known to {\em Company Y} are placed in protected form on an EaaS infrastructure, and some examples of these data are made public. People participate by submitting VMs containing their executable software to the EaaS infrastructure, where the VMs are sandboxed and the executables are run on the data. The standardised VM exchange format reduces the overhead for participants. Once the VM has run and produced results, it is destroyed (to ensure privacy of the data), while the outputs of the executables remain available for {\em Company Y} and the evaluation metric values are returned to the participant that submitted the VM. Participants can choose to make these metric values public, and receive a Digital Object Identifier allowing these results to be referred to in a publication. Visual analytics software on the EaaS infrastructure allows participants a detailed analysis of the performance of their algorithms and a comparison to other publicly available results. {\em Company Y} can examine combinations of the outputs of the submitted software of the two parts of the challenge to find the optimal combination, have experts examine newly extracted facts or predictions to evaluate their relevance (increasing the size of the ground truth), and contact the participants having the best performing submissions to negotiate terms for the further use and development of their techniques.
\end{quotation}

We now outline what we see as the main contributions of the full EaaS paradigm, in terms of benefits, simplicity, reproducibility and cooperation. We also consider the points of view of both academia and industry.
\subsection{Benefits}
In designing and implementing the EaaS paradigm, it should be ensured that the benefits significantly outweigh the effort invested both from academia and industry. One of the main benefits to both sides is that the EaaS paradigm should bridge the gap between academia and industry, allowing more straightforward cooperation. This would in particular counter the commonly perceived view that scientists working in industry have access to larger amounts of more interesting data than those working in academia~\cite{Hub12,Mar12}. The EaaS paradigm benefits researchers in academia because it allows them access to industry data, but caters to industry because there is no necessity to release the data in any uncontrolled way, as the data can remain behind a firewall on company servers. Furthermore, through making available data and associated challenges through the EaaS paradigm, companies get exposure to the latest processing and analysis approaches from academia. The EaaS paradigm also makes A/B testing~\cite{Kohavi2012} on real-time data available to researchers in academia, which again brings benefits to companies owning the data, as they get results on dynamic data currently of interest to them, rather than on static data from months or years back (the sort of data that could be considered less commercially interesting and hence more suitable to release to researchers in the traditional way).

For researchers in academia, beyond the obvious benefits of access to large amounts of interesting data linked to challenges of commercial relevance, there are also potential benefits in terms of increased reputation for organisers and participants. This would be particularly true for those EaaS instances that become accepted as benchmarks in a scientific area rather than one-off competitions. In a fully developed EaaS infrastructure, participants could also have access to technical benefits, such as an interface for performing visual analytics of the experimental results to potentially even a service to semi-automatically write the experimental section of a scientific paper. Potentially, submissions could also be automatically encapsulated as services to be made available on a demo webpage, thereby also increasing the visibility of the participants' work.

Both academia and industry can take advantage of the capability inherent in submitting VMs containing functioning services, namely the running of automated ensemble approaches. It is well known that ensembles of classifiers can often perform better than a single classifier~\cite{Die00} --- the EaaS paradigm makes it straightforward to test multiple combinations of classifiers in various ensemble approaches to obtain new scientific insight as well as better performing classifiers.

Finally, EaaS instances can be used as part of university courses on Data Science to give students experience with working on real challenges on huge amounts of data, instead of the ``toy problems'' that are generally part of their course work. This brings a benefit to industry by ensuring that university graduates entering the Data Science job market are better qualified for the work that they will have to do. Such EaaS instances could, for example, be combined with a Coursera course.
\subsection{Simplicity and Cost-Effectiveness}
The benefits of EaaS should significantly outweigh the effort required from all stakeholders. While the previous section concentrated on making the benefits explicit, this section focuses on how the effort can be reduced. 

To reduce the effort needed from organisers of EaaS instances, they should not be forced to set up the full infrastructure necessary for EaaS each time they organise a competition or benchmark, as this would be an unacceptable overhead. Optimally, EaaS infrastructures should be available that can be used for reasonable costs. These infrastructures should also be easily scalable in terms of the number of participants, so that the effort for a small and large number of participants is similar. Furthermore, there should be effective support in carrying out all the steps required in setting up the challenge, including steps such as designing an effective evaluation protocol and selecting the most suitable metrics for the task.

The effort required for participants must also be reduced. An effective way of doing this is to use a common VM format that can be executed on all EaaS infrastructures. This means that participants could have prepared VMs containing their algorithms, and can submit them easily to a benchmark or competition on any EaaS infrastructure, avoiding the need to spend time in adapting the code to multiple cloud architectures. Where possible, standardisation of other aspects of a submission, such as data formats, could also lower participation overheads.
\subsection{Reproducibility}
The drawbacks of publishing papers containing the results of experiments done only on proprietary data that is not available to other researchers to ensure reproducibility of results has been widely discussed~\cite{CM12}. The EaaS paradigm can contribute to increasing reproducibility of results through making available not only data and associated tasks over a long term, but also a library of executable algorithms that have been applied to solving the tasks on the data, and the results that have been produced by these algorithms. 

The availability of these resources should contribute to addressing an observed practice in computer science of comparing new algorithms to weak baselines~\cite{AMW2009a}. It should also ensure access to results using a large palette of metrics, so that all aspects of the performance of an approach can be examined. This can include execution time info, as every approach is evaluated on the same infrastructure.  

Further contributions toward reproducibility can also be expected. As the dataset is stored centrally, mechanisms can be put in place to collect additional ground truth for a task over time. Crowd-sourcing, either among competitors or among a wider group of people, could be used to obtain this additional ground truth~\cite{FoM2012}. Whenever the ground truth available has expanded significantly, all approaches already submitted for a task could be re-evaluated automatically using the expanded ground truth, which then becomes the standard for future submissions. Through the use of a publishing approach such as executable papers, it can be ensured that the latest results for a published approach are always available. 

The EaaS setup would also allow research that is not easily possible today. An example is the development of new performance metrics better adapted to modelling specific tasks. Researchers developing new metrics would be able to experiment at a large scale on how the use of a new performance metric affects the ranking of submitted approaches in comparison to existing metrics.
\subsection{Cooperation}
EaaS could also lead to new ways to encourage cooperation and form multi-disciplinary teams. For tasks requiring a collection of complementary skills to solve them, the insight provided by EaaS results into how different types of approaches perform could assist in effective team formation. This also means that participating teams have the possibility to concentrate on the aspect of the task for which they feel most qualified, and collaborate with other participants (either explicitly or through re-use of their approaches) to cover those aspects of the task for which they have lower expertise.  
\section{Advancing EaaS}
\label{sec_advancing}
This section presents the next steps in advancing EaaS in terms of the technical, acceptance and regulatory aspects, which were identified as the key aspects of EaaS.
\subsection{Technical Aspects}
There is more than one way to implement an Evaluation-as-a-Service platform, and it is as of yet unclear which way is the best one or which of the ways that have been pursued so far will prevail. We review design choices for Evaluation-as-a-Service platforms from the perspective of organizers and participants in a shared task competition, as well as with regard to reproducibility properties of the evaluation results obtained; both along the three technical aspects ``Implementation,'' ``Submission,'' and ``Execution.'' Table~\ref{table-eaas-technical-aspects} gives an overview.
\subsubsection{Implementation}
The implementation aspect refers to the programming efforts that are expected from participants. Basically, four alternatives can be distinguished:
\begin{description}
\item[Software:]
Participants are asked to implement their entire software themselves without any restrictions regarding programming languages used, its architecture, its components, or its interface. The only restriction is given by the format of the input data, and the expected format of its output as defined by the organizers. This is the default modus operandi of almost all shared task competitions to date, so that it serves as baseline for this aspect.
\item[Plugins:]
Participants are presented with a fully-fledged piece of software that features a plugin architecture, where a plugin is supposed to solve the problem underlying the shared task. In this case, the programming language and the interfaces are pre-defined and cannot be chosen by participants. A plugin architecture also usually prescribes by which approaches a given problem can be solved, or at least limits the space of possible approaches at solving the shared task's underlying problem.
\item[Modules:]
Participants are asked to implement a software module that is integrated in a given processing pipeline. The organizers of a shared task competition have to specify the software architecture of the pipeline up front, specifying the interfaces of all modules. Moreover, organizers need to provide baseline implementations of all modules up front. Participants may then choose which of the modules they wish to replace with their own implementations, but are at liberty to resort to the baseline implementations. Restrictions to programming languages may be avoided in this case if the module interface is, for example, a POSIX command line interface with pre-specified input and output formats for each module.
\item[Services:]
Participants are asked to implement their software as a web service with a pre-specified API. In this case, no restrictions apply with regard to how the service is implemented internally, whereas implementing and hosting a web service creates an overhead for participants.
\end{description}
When considering Plugins, Modules, and Services as alternatives to the default, Software, each one brings about its own pros and cons. In general, it can be said that all of these alternatives have negative side effects on participants and organizers alike, but choosing them may improve the long-term effects of a shared task competition significantly (see Table~\ref{table-eaas-technical-aspects}). 

The flexibility of participants is much worse with Plugins, whereas Modules, and particularly Services conserve some flexibility. Setting up Services require much more effort from participant than developing software in isolation, whereas Plugins and especially Modules require less effort, respectively. From the organizer point of view, up front effort and ongoing effort are worst when asking for Plugins, because the entire software has to be provided in a working condition to participants, who will argue a lot about details. That is less so, when asking for Modules or Services, since here only some up front effort is required, mostly specifying interfaces. Little to no extra infrastructure is required with either solution, disregarding software provided by organizers, but Plugins require significant person hours from organizers, whereas Modules less so, and Services even less.

Regarding evaluation results, their repeatability, reproducibility, sustainability, and efficiency will be much better when employing one of the alternatives to Software. There is one exception, though, namely Services: one cannot rely on participants of a shared task competition to host and maintain their services for a long time after the competition has passed. Therefore, this alternative is much less sustainable and its repeatability and reproducibility depends on whether a participant's service is still available. The service can be employed on a central infrastructure though, as well.
\input{table-eaas-technical-aspects}
%
\subsubsection{Submission}
The submission aspect refers to what piece of data participants are supposed to submit to a shared task competition. Basically, three alternatives can be distinguished:
\begin{description}
\item[Run Output:]
Participants are asked to submit the output of their software when it is executed on a test dataset supplied by the organizers. The output's format has to comply with a pre-specified format supplied them. This is the default modus operandi of almost all shared task competitions to date, so that it serves as baseline for this aspect.
\item[Source Code:]
Participants are asked to submit the source code of their software, whereas the software must comply with the input and output formats specified by the organizers. Furthermore, participants are asked to supply instructions as to how to get the source code compiled and running.
\item[Compiled Binary:]
Participants are asked to submit compiled binaries of their software in the form libraries or executables. Moreover, participants must supply instructions about dependent software as well as how the binary can be executed.
\end{description}
When considering the submission of Source Code and Compiled Binaries as altenative to the default, submitting Run Output, the former obviously require much more effort from organizers (e.g., ongoing, infrastructure, and man hours), whereas participants have little to no overhead besides providing documentation (see Table~\ref{table-eaas-technical-aspects}). However, it is just as obvious that submitting Source code or at least Compiled binaries has a significant benefit in improving the long-term effects of shared tasks in terms of repeatability, reproducibility, and sustainability.
\subsubsection{Execution}
The execution aspect refers to where the participant software is executed in order to obtain its output for evaluation. Basically, three alternatives can be distinguished:
\begin{description}
\item[Local:]
Participants are asked to execute the software locally using their own hardware. For this purpose, organizers must provide test datasets to participants, typically without revealing the ground truth. This is the default modus operandi of almost all shared task competitions to date, so that it serves as baseline for this aspect.
\item[Managed:]
Organizers execute the participant software on their own hardware. This presumes that participants submit either source code or a compiled binary. This way, organizers need not release test datasets to participants.
\item[Virtualized:]
Participants are asked to deploy and execute their software in a virtual machine provided by the organizers. In this case, it depends on whether the software execution can be handled remotely by participants to decide whether the test datasets need be directly accessible to participants to execute their software, or not.
\end{description}
When considering the Managed or Virtualized execution of participant software as alternative to the default, Local execution, there are negative side effects for both participants and organizers. Organizers face the problem of executing untrusted software, and they need to provide the infrastructure for a timely execution of submissions. Participants may prefer a Virtualized execution over Managed execution, since the former gives them more control over their software, and whether it works as advertised, whereas the latter has a high turn-around time for them. Again, a significant improvement for the repeatability, reproducibility, sustainability, and even efficiency may be attained when choosing one of the alternatives to Local execution.
\subsubsection{Automation Avoids Negative Side Effects}
To make any form of Evaluation-as-a-Service viable, one must depart from the default approaches to Implementation, Submission, and Execution. However, doing so incurs significant risks for organizers both in terms of time spent as well as driving away participants. The aforementioned initiatives have still taken these risks, and managed them by some means of automation. In general, assisting participants and organizers with automating as many of the aforementioned technical aspects as possible will increase the acceptance of Evaluation-as-a-Service, ideally reducing its overhead to a point at which going with the default options is attractive. Some aspects, however, cannot be automated, however, these concern only the organizers of a shared task competition: for example, a Plugin architecture has to be built on a task-by-task basis, and only certain components of such an architecture may be shared across tasks. Nevertheless, the successes of the aforementioned initiatives in prototyping Evaluation-as-a-Service platforms suggest that, in time, the development of robust automated services to assist participants and organizers in adopting this paradigm will become available.
\subsection{Acceptance Aspects}
Besides mastering the technical challenges, one further key step for advancing EaaS is to communicate the benefits of the paradigm to the research community, funding agencies and companies and to overcome expressed or experienced concerns. 
On the one hand this can be achieved by providing compelling incentives for participation, on the other hand by countering the fears the various stakeholders of EaaS campaigns might have.
Lowering the entry barriers would be an important first step, so something simpler than virtual machines or running code, but something easy to replicate from a local installation. Docker containers might be a solution but they need to become available for several platforms (Windows and MacOS potentially as well) and also in security models suitable to run them in a variety of contexts.

To motivate participation it is also important to look at which type of participants a challenge aims at, for example those interested in prize money, publications or free T-shirts and then address the desired community very clearly.

Sustainability of infrastructures including data availability and computing needs to be assured and not only linked to short-term projects. Likely this will require public-private partnerships that need to be based both on academic and industrial needs and aim at the long term and not only a short period to be sure to benefit from the main advantages.

Funding agencies can help by engaging in campaigns and infrastructures as they can have high benefit but need to motivate funded groups to not only make data available but engage in serious performance comparison and collaboration with other groups. Data science has many challenges in terms of managing the data and keeping research data available. 
This should likely rather highlight a collaboration aspect and not a pure competition that often does not offer incentives for collaborations.

Trust needs to be built as well that all challenges are objective and that no cheating is possible and each participant has the same chances. Also, for company participants that would like their code protected or data providers such as hospitals that need to assure that no data leaves an infrastructure, trust is an essential part. Such trust can likely only be built over time and with longer term experiences and this is what the systems should be optimized for.

Another strong part of acceptance is the creation of a community feeling that involves all partners from data and problem providers to participants and that takes their comments into account. Also the communication between participants can be fostered in this way, lowering entry barriers and increasing the collaboration that can benefit all participants.

Standardization across tools and data is another point that can help acceptance and this could be a top-down problem as the diversity in research is high and bottom-up standardization might be harder to achieve. Standardization can be related to data formats, interfaces of tools and components and portability of the software containers.

In the end, for all partners the personal benefits are an important criterion and these need to be made visible and measured to increase motivation. This can be the case for both participants and providers of tasks and challenges. There should be local optimization on the researcher level but particularly much more global optimization.

In the long run such EaaS has to be integrated into the entire research process and similar to current initiatives for data sharing by funding agencies, clear financial incentives in the entire research process can help everyone involved. Such integration needs to be done on an international level beyond current national or regional funding bodies if possible. 
\subsection{Regulatory Aspects}
%
This section discusses the regulatory aspects of EaaS from two viewpoints: legal considerations for effectively running EaaS competitions, and potential steps toward running EaaS sustainably.
\subsubsection{Legal Considerations}
\begin{figure}[tb]
\begin{center}
\includegraphics[width=\textwidth]{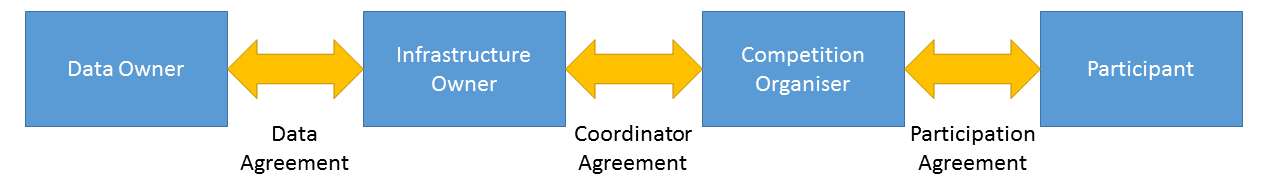}
\caption{The stakeholders and potential agreements necessary to organise an EaaS competition.}\label{fig_agreements}
\end{center}
\vspace{-4ex}
\end{figure}

Four groups of stakeholders were identified in the organisation of an EaaS competition:
\begin{description}
\item[Data Owner:]{The organisation that owns and provides the data to be used in the EaaS competition. This could be one organisation or a group of organisations for a more complex task, which can also include distributed data storage and execution.}
\item[Competition Organiser:]{The organisation or group of organisations that define the tasks to be solved on the data for the competition, specify the evaluation criteria and administer the EaaS competition.}
\item[Infrastructure Owner:]{The organisation or group of organisations providing the infrastructure for running the EaaS competition. There could be more than one organisation involved if, e.g., one organisation provides the infrastructure on which the EaaS competition is run, while another provides the software to administrate the EaaS competition.}
\item[Participant:]{The organisations or individual people participating in the EaaS competitions.}
\end{description}
The following three levels of necessary legal regulation were identified for EaaS. For each of these levels, the stakeholders involved are mentioned, and a diagram showing the relation between the stakeholders and the agreements is shown in Figure~\ref{fig_agreements}. 
\begin{description}
\item[Data:]{This includes aspects such as regulating the appropriate use of the data, ensuring consistency in the terms of data release, and certification of an infrastructure to host a specific data type (e.g., HIPAA). The {\em Data Owner} and {\em Infrastructure Owner} stakeholder groups are involved in this agreement.}
\item[Participation:]{This includes the rules for participation in a specific EaaS competition, regulating, e.g., withdrawal from participation and permitted channels of result publication (such as no use of results in advertisements). The {\em Competition Organiser} and {\em Participant} stakeholder groups are involved in this agreement.}
\item[Coordinators:]{This regulates what the coordinators may and may not do, including the re-use of programs submitted by participants on further data. The {\em Competition Organiser} and {\em Infrastructure Owner} stakeholder groups are involved in this agreement.}
\end{description}
In order to facilitate the organisation of an EaaS competition, standardised templates for these three agreements would be useful. Optimally, it would be possible to automatically generate the agreements based on options selected on a website, although this is made more complex as different agreements would be needed for different jurisdictions. These agreements should also take into account various specific requirements by organisations, such as the possibility for a participating company to embargo results, specific data requirements of some government research laboratories, and foreseeing the use of Non-Disclosure Agreements in some cases. Even if these agreements do exist, there is the complication of the enforceability of participant agreements signed in other countries. A clear chain of liabilities will also have to be defined.

In order to make the organisation of EaaS competitions as straightforward as possible and avoid extremely complex legal agreements, a set of guidelines covering the best case of organisation should be released. This would include suggestions such as the following:
\begin{itemize}
\item data needs to be released under conditions that allow it to be as broadly usable as possible;
\item non-anonymous data should only be used on a secure infrastructrue, but this still involves some risk;
\item the algorithm creator should agree to the broadest possible terms, in the best case an open source release (or at least making the code available), and allowing use of the submitted algorithms on further data at the discretion of the organisers.
\end{itemize}
\subsubsection{Sustainability}
EaaS has an additional cost beyond standard evaluation campaigns and competitions in that it needs an infrastructure on which to run the EaaS. It therefore needs to provide a clear return on investment for a company to organise such a competition. Two potential sources of return on investment are identified here:
\begin{description}
\item[Open Innovation:]{Through making challenging tasks available as competitions, companies can receive potential solutions to their challenges from a significantly larger number of experts than would be available within the company. For this to work, participants have to agree to conditions for the company to continue to use their work (e.g., in the participation agreement).}
\item[Access to Talent:]{Companies could hire the people providing the best solutions to the challenges, therefore getting access to the best matches in terms of skill. This could also be used by venture capitalists to identify talent to fast-track to a new incubator.}
\end{description}
Examples of successes with the EaaS paradigm are given in Section~\ref{sec_successes} of this white paper. Due to the impact that EaaS can have on innovation, it would also make economic sense for an EaaS infrastructure to be supported by public funds, at least in an initial stage until a business model for running competitions on behalf of companies and other organisations can be put in place as a public-private partnership.
%
%
\section{Conclusions}
\label{sec_conclusion}
%
Evaluation campaigns have advanced many scientific areas and fields and focused research also in economic areas via platforms such as Kaggle that proposes machine learning challenges. Several companies have managed to make a business out of these and crowdsourcing part of the machine learning development can benefit in many areas to obtain and use optimized solutions.
The impact is important and has advantages for organizers of such challenges, but also to participants, and companies who can propose their research challenges.

Evaluation-as-a-Service was created due to problems with the typical challenge of distributing large test data sets, working with confidential data that cannot be shared, and real time data that cannot be packaged. Several approaches have been created over the past few years to respond to the shortcomings, and different solutions were developed that are compared in this white paper. 
The white paper was started at a workshop on EaaS in March 2015 in Sierre, Switzerland, but has evolved since then and become much more concrete with many aspects being detailed based on experiences.

EaaS has the potential to change the way challenges are run and to integrate with other initiatives such as clouds in the scientific sector to create more efficient and effective research infrastructures in the future.
Motivations are manifold, both for funding agencies, organizations proposing data and tasks for challenges but also challenge organizers and participants in terms of impact and best use of available funding.

It is foreseen that EaaS will, once it is further developed, ensure reproducibility from citable data to executable papers and the possibility to run existing tools on new data directly to create strong baselines automatically and assess the best techniques. It will make routine tasks automatic and concentrate real effort on novelty and improving existing techniques. Common platforms should also foster experience sharing and comparison of components, something that has not always been successful in past challenges. It can be much easier with central data and all tools accessing this data on the same platform, as has been done in some very specific domains, for example with NTRIC.\footnote{\texttt{https://www.nitrc.org/}}

Big data and data science need new approaches to create a sustainable research infrastructure and we expect EaaS to be a central part of such an infrastructure. Particularly the ever-increasing amount of data created and analysed creates challenges that are not easy to resolve
.
Many challenges still need to be addressed but much experience has already been gained via the existing approaches and this creates a solid foundation for the next steps.
\section*{Acknowledgements}
We acknowledge financial support by the European Science Foundation via its Research Network Program ``Evaluating Information Access Systems'' (ELIAS) and by the European Commission via the FP7 project VISCERAL (318068).
\bibliographystyle{plain}
\bibliography{eaas}
\end{document}

%% file: table-eaas-initiative-comparison.tex
\begin{table}%
\centering%
\footnotesize%
\renewcommand{\tabcolsep}{1pt}%
\renewcommand{\arraystretch}{1.1}%
\caption{Comparison of selected EaaS initiatives}%
\label{table-eaas-initiative-comparison}%
\vspace{2ex}%
\begin{tabulary}{\linewidth}{@{}LLLLLLLLL@{}}
\toprule
\textbf{Initiative} & \textbf{Software} & \textbf{Data} & \textbf{Data Access} & \textbf{Submission} & \textbf{Continuous} & \textbf{Automation} & \textbf{Result Interaction} & \textbf{Technical Support} \\
\toprule
TREC Microblog 2013-2014 & Twitter Tools & Static & API   & Result file upload & Fixed deadline & Little & None  & Online forum \\
\midrule
BioASQ & Dedicated & Static / Dynamic & Download & Result file upload & Fixed deadline & Part & Online leaderboard & Online forum \\
\midrule
VISCERAL Anatomy1/2 & VISCERAL Registration System & Static & VM    & VM    & Fixed deadline & Little & None  & Mailing list \\
\midrule
VISCERAL Anatomy3 & VISCERAL Registration System & Static & VM    & VM    & Continuous & Full  & Online leaderboard & Mailing list \\
\midrule
CLEF NewsREEL & Open Recommendation Platform & Dynamic & API   & API   & Fixed deadline & Part & Online leaderboard  & Tutorials \& Mailing list\\
\midrule
CLEF LL4IR & Living Labs API & Static & API / Download & API / Upload & Fixed deadline & Part & None  & Mailing list \\
\midrule
C-BIBOP & Codalab & Static & Download, API planned  & Result file upload, code upload, Docker planned   & Fixed deadline and continuous   & Full   & Online leaderboard   & Online forum \\
\midrule
PAN Evaluation Lab & TIRA  & Static & VM & VM & Fixed deadline & Full  & Web front end & Mailing list \\
\midrule
CoNLL Shared Task 2015 & TIRA  & Static & VM & VM & Fixed deadline & Full & Web front end & Mailing list \\
\midrule
TREC Total Recall Track & Baseline Model Implementation & Static & API / Download & VM / Script & Fixed deadline & Part   & None  & Online forum \\
\midrule
MIREX & MIREX submission system & Static & None  & Compiled Code  & Fixed deadline & Little & None  & Mailing List \\
\bottomrule
\end{tabulary}%
\end{table}

%% file: table-eaas-technical-aspects.tex
\providecommand{\worse}{{\bfseries{}--}}
\providecommand{\Worse}{{\bfseries{}--~--}}
\providecommand{\WORSE}{{\bfseries{}--~--~--}}
\providecommand{\better}{+}
\providecommand{\Better}{+~+}
\providecommand{\BETTER}{+~+~+}
\providecommand{\same}{\Circle}
\providecommand{\both}{+~{\bfseries{}--}}

\begin{table}%
\centering%
\footnotesize%
\renewcommand{\tabcolsep}{2.25pt}%
\renewcommand{\arraystretch}{0.9}%
\caption{Comparison of alternative implementation approaches for the Evaluation as a Service paradigm. We distinguish implementation from submission and execution of participants software. The traditional approach to organizing shared task competitions serves as a baseline in each case (top row of each category). The alternative approaches are judged whether they perform much worse (\WORSE), worse \mbox{(\Worse)}, a little worse (\worse), similar (\same), a little better (\better), better (\Better), or much better (\BETTER) compared to their respective  baseline with respect to 10 criteria (columns). For some criteria, an approach performs better or worse (\both), dependent on the circumstances.}%
\label{table-eaas-technical-aspects}%
\vspace{2ex}%
\begin{tabular}{@{}l@{\kern-1em}cccccccccc@{}}
\toprule
\multirow{2}{5em}{\bfseries EaaS aspect\vspace{-2pt}} & \multicolumn{2}{@{}c@{}}{\bfseries Participant} & \multicolumn{4}{@{}c@{}}{\bfseries Organizational effort} & \multicolumn{4}{@{}c@{}}{\bfseries Evaluation result} \\
\cmidrule(r@{\tabcolsep}){2-3}\cmidrule(l@{\tabcolsep}r@{\tabcolsep}){4-7}\cmidrule(l@{\tabcolsep}){8-11}
& flexibility & effort & up front & ongoing & infrastructure & man hours & repetition & reproduction & sustainment & efficiency \\
\midrule
\multicolumn{2}{@{}l@{}}{\em Implementation} \\
\midrule
Software        & \same   & \same   & \same   & \same   & \same   & \same   & \same   & \same   & \same   & \same   \\
Plugins         & \WORSE  & \Worse  & \WORSE  & \Worse  & \worse  & \WORSE  & \BETTER & \BETTER & \BETTER & \BETTER \\
Modules         & \Worse  & \worse  & \Worse  & \worse  & \Worse  & \Worse  & \BETTER & \Better & \BETTER & \Better \\
Services        & \worse  & \WORSE  & \Worse  & \worse  & \same   & \worse  & \both   & \both   & \WORSE  & \better \\
\midrule
\multicolumn{2}{@{}l@{}}{\em Submission} \\
\midrule
Run output      & \same   & \same   & \same   & \same   & \same   & \same   & \same   & \same   & \same   & \same   \\
Source code     & \same   & \Worse  & \same   & \WORSE  & \WORSE  & \WORSE  & \BETTER & \BETTER & \BETTER & \same   \\
Compiled bin.   & \same   & \worse  & \same   & \Worse  & \WORSE  & \Worse  & \BETTER & \Better & \Better & \same   \\
\midrule
\multicolumn{2}{@{}l@{}}{\em Execution} \\
\midrule
Local           & \same   & \same   & \same   & \same   & \same   & \same   & \same   & \same   & \same   & \same   \\
Managed         & \WORSE  & \Worse  & \worse  & \WORSE  & \WORSE  & \WORSE  & \BETTER & \Better & \BETTER & \BETTER \\
Virtualized     & \worse  & \worse  & \WORSE  & \Worse  & \WORSE  & \WORSE  & \BETTER & \BETTER & \BETTER & \Better \\
\bottomrule
\end{tabular}
\end{table}
